\documentclass[a4paper]{article}
\usepackage{url}
\usepackage{multirow}
\usepackage{INTERSPEECH2021}
\usepackage{cite}
\usepackage{tablefootnote}
\usepackage{color}
\usepackage{booktabs}
\usepackage{graphicx} 
\usepackage{epstopdf, multirow}
\usepackage{pifont}
\usepackage{footnote}
\usepackage{soul}
\usepackage{amssymb}
\usepackage{pifont}
\newcommand{\textoverline}[1]{$\overline{\mbox{#1}}$}
\usepackage{todonotes}
\title{Unsupervised Acoustic Unit Discovery by Leveraging a Language-Independent Subword Discriminative Feature Representation}

\name{Siyuan Feng$^{1}$,
Piotr \.Zelasko$^{2,3}$, 
Laureano Moro{-}Vel{\'{a}}zquez$^{2}$ and
Odette Scharenborg$^1$
\thanks{Code: \texttt{https://github.com/syfengcuhk/mboshi}. We thank Lucas Ondel for valuable discussion on the evaluation software.}
}

\address{
  $^{1}$Multimedia Computing Group, 
  Delft University of Technology, Delft,  the Netherlands
  $^{2}$Center for Language and Speech Processing, $^{3}$Human Language Technology Center of Excellence, \\Johns Hopkins University, Baltimore, MD, USA
  }
\email{\{S.Feng, O.E.Scharenborg\}@tudelft.nl, \{petezor,laureano\}@jhu.edu}

\begin{document}

\maketitle
\begin{abstract}
This paper tackles  automatically discovering  phone-like acoustic units (AUD) from unlabeled speech data. 
Past studies usually proposed single-step approaches.
We propose a two-stage approach: the first stage learns a subword-discriminative feature representation, and the second stage applies clustering to the learned representation and obtains phone-like clusters as the discovered acoustic units. In the first stage, a recently proposed  method in the task of unsupervised subword modeling is improved by replacing a monolingual out-of-domain (OOD) ASR system with a multilingual one to create a  subword-discriminative representation that is more language-independent. In the second stage, segment-level k-means is adopted, and two methods to represent the variable-length speech segments as fixed-dimension feature vectors are compared. Experiments on a very low-resource Mboshi language corpus   show that our approach outperforms state-of-the-art AUD in both normalized mutual information (NMI) and F-score. The multilingual ASR improved upon the monolingual ASR in providing OOD phone labels and in estimating the phone boundaries. 
A comparison of our systems with and without knowing the ground-truth phone boundaries showed a 16\% NMI performance gap, suggesting that 
the current approach can significantly benefit from improved phone boundary estimation.
 \end{abstract}

\section{Introduction}
There are around $7,000$ spoken languages in the world \cite{austin2011cambridge}, most of which lack transcribed speech data \cite{adda2016breaking}. Conventional supervised acoustic modeling strategies \cite{dahl2011context,chan2016listen} therefore cannot be applied directly to build ASR systems for such low-resource languages. As a result, current high-performance ASR schemes are available only for a very small number of languages \cite{feng2021how}.
To facilitate ASR for low-resource languages,  unsupervised acoustic modeling  has been gaining research interest  recently \cite{lee2012a,I3EWang,chen2015parallel}.  Unsupervised acoustic modeling aims to discover basic speech units that represent  all the sounds in a target language by making a  \textit{zero-resource} assumption  \cite{dunbar2017zero}, i.e., for a target language, only speech recordings are available while transcriptions and phoneme inventory (or its size) information are unknown. 

There are two mainstream research strands in unsupervised acoustic modeling. The first strand, \textit{acoustic unit discovery (AUD)} \cite{ondel2016variational,lee2012a}, formulates the problem as discovering a finite set of phone-like acoustic units \cite{lee2012a,I3EWang,Ondel2019Bayesian}. 
The second strand, \textit{unsupervised subword modeling (USM)} \cite{dunbar2017zero,Dunbar2019}, formulates the problem as learning a frame-level feature representation  that can distinguish subword units (phonemes)  and is robust to speaker variation \cite{chen2015parallel,oord2017neural,heck2017feature}. 
Studies on the USM task were mostly driven by the 
ZeroSpeech Challenges
\cite{dunbar2017zero,Dunbar2019,Dunbar2020zero}. 
In essence, the USM task can be considered as learning an intermediate representation towards achieving the goal of AUD \cite{Feng2019combining}. 

This study addresses the AUD task. 
Two main types of approaches to  the AUD task were investigated in the past. The first type adopts  self-supervised  learning algorithms and uses a quantization layer to obtain a finite set of discovered acoustic units \cite{oord2017neural,baevski2020vqwav2vec,Niekerk2020vector}. The second type adopts Bayesian non-parametric versions of the hidden Markov model (HMM) \cite{lee2012a,Ondel2019Bayesian,Yusuf2020hierarchical}. The combination of self-supervised learning and Bayesian approaches was also studied \cite{Ebbers2017,ondel2018bayesian}.
All the studies mentioned above proposed single-step approaches. In contrast, the present study proposes a two-stage learning framework: the first stage learns a frame-level subword-discriminative  feature representation  (i.e., the USM task); the second stage  applies clustering techniques to the  learned frame representation to obtain a set of clusters as the discovered acoustic units. 
Subword-discriminative feature representations can provide a better separation between sounds than spectral features: In a subword-discriminative representation,  two   examples of the same phoneme are closer while those of different phonemes are further apart than in an  MFCC representation. This is a    highly desired property in     clustering-based acoustic unit discovery \cite{I3EWang,Bhati2019unsupervised}, which motivates us to propose a two-stage learning framework.



Specifically, in the first stage of the framework proposed in this study, we leveraged and improved a USM approach from a previous study \cite{feng2020unsupervised}. This approach trains an autoregressive predictive coding (APC)  model \cite{Chung2019} followed by a cross-lingual DNN model to  extract bottleneck features  (BNFs)  as the   subword-discriminative representation. 
Previous results  \cite{feng2020unsupervised,feng2020effectiveness} employing a \textit{single} out-of-domain (OOD) language's resources to generate OOD phone labels for cross-lingual DNN training,  provided state-of-the-art performance in USM tasks. 
Here, we aim to improve this approach further and, for the first time, use it for a different task: AUD. 
To leverage an OOD ASR system to generate OOD phone labels, We propose to use \textit{multiple} OOD languages' resources to build a more language-independent OOD ASR system than in our previous work \cite{feng2020unsupervised}. 
Because different languages have different phoneme inventories, we hypothesize that OOD phone labels that capture a more extensive set of sounds will be more useful for acoustic modeling of a target language. We will compare the use of multiple versus a single OOD language resources in this paper.
In the second stage of our framework,  the $k$-means algorithm is adopted for speech segment clustering. To that end, first, phone segment boundaries are estimated using an OOD ASR via decoding \cite{feng2016exploit}. The resulting variable-length segments need to be represented as fixed-dimension vectors, 
for which we compare two often adopted approaches \cite{kamper2017embeded,Bhati2019unsupervised,feng2016exploit}: an average-based method \cite{I3EWang} and a downsampling method \cite{levin2013fixed}.  
We measure the sensitivity of our AUD approach's performance to   the number of discovered acoustic units,
because the phone inventory size of a target language is usually unknown. 
The experiments   are carried out   on a very low-resource language, Mboshi  \cite{Godard2018mboshi} ($4.5$ hours of unlabeled speech).



\section{Proposed two-stage approach}
\subsection{Stage 1: subword-discriminative feature learning}
\label{subsec:approach_1st}
\begin{figure}[!t]
    \centering
    \includegraphics[width =  \linewidth]{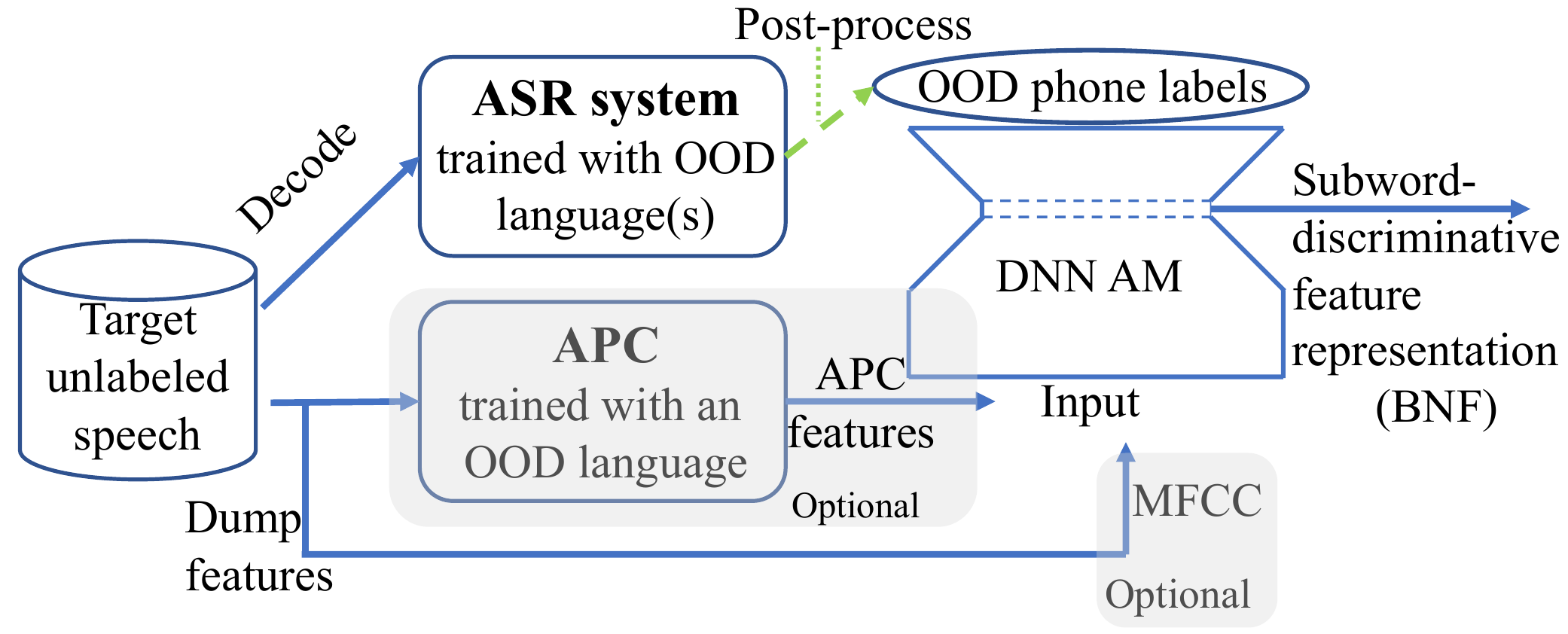}
    \caption{The first stage in the proposed approach. ``OOD'' denotes out-of-domain, i.e. non-target language(s). The input to the DNN AM is either APC features or MFCCs but not both.}
    \label{fig:framework_1st}
\end{figure}

The goal of this stage is to learn a frame-level subword-discriminative feature representation.
The  general framework of the first stage of our proposed approach, based on \cite{feng2020unsupervised}, consists of an 1) APC model, which creates the APC features that are used as input to a  2) DNN acoustic model (AM) together with the frame-level phone labels obtained from 3) an OOD (non-target language)  ASR system
(see Figure \ref{fig:framework_1st}). After training the DNN AM, BNFs are extracted from the bottleneck layer of the DNN as the desired subword-discriminative representation.

APC is a self-supervised learning model without the need of transcriptions for training. It is trained to predict a future speech frame $n$ steps ahead (named \textit{prediction step}) based on the current and past frames of an utterance \cite{Chung2019}. APC is incorporated in \cite{feng2020unsupervised} to extract \textit{APC features} as input to the DNN AM (see Figure \ref{fig:framework_1st}), owing to its ability to make phonetic and speaker information   in speech   more separable than  MFCC.  
In this study we tackle an extremely low-resource scenario ($4.5$ h), however
our previous work \cite{feng2020unsupervised} suggests that a larger training dataset (more than $50$ h) is needed to obtain effective APC features, thus we opt  for using an APC model trained with a well-resourced OOD language. Note that a recent study \cite{riviere2020unsupervised} found that self-supervised models trained on one language can be used to represent another language with certain success. In our experiments we compare (1)  APC features extracted by an APC model which is trained with an OOD language; and (2) MFCC features (hence bypassing the APC model); as the input  to the DNN AM (see Figure \ref{fig:framework_1st}). 


The  DNN AM in Figure \ref{fig:framework_1st}
is trained with target language acoustic data. 
Frame-level phone labels required for training the DNN AM are obtained using an OOD (non-target)  ASR system  \cite{feng2020unsupervised}: a  target speech utterance is decoded by the OOD ASR system so that every frame is assigned  a  phone label generated by the OOD ASR. 
By this means, an OOD language's phonetic knowledge is exploited for the target language acoustic modeling.
BNFs are then extracted from the bottleneck layer of the trained DNN as the desired subword-discriminative representation.


The language-independent OOD ASR system leverages multiple phonetically diverse languages' resources. We use International Phone Alphabet (IPA) symbols \cite{international1999handbook}   to represent the phonemes of the different OOD languages, thus creating a phoneme inventory of the multilingual ASR system that is more language-independent \cite{zelasko2020sounds} than that of a monolingual OOD. This also enables acoustic information sharing of  the
same or similar 
sounds from multiple  languages during   ASR  training.   A multilingual ASR system captures a wider phonetic space and has more different phone labels than a monolingual ASR system, thus is expected to provide more refined OOD phone labels for the target speech than a monolingual  ASR system. 

Another modification to the approach in \cite{feng2020unsupervised} is adding a post-processing step to  the OOD ASR based phone labels (see Figure \ref{fig:framework_1st}). Essentially, the post-processing aims to refine the phone labels from an OOD ASR via re-aligning:
First, we train an HMM with the OOD phone labels and Mboshi acoustic data; second, we generate the HMM phone alignments as the  desired frame-level label supervision (rather than the output of an OOD ASR) to train the DNN AM. We  experimentally found that the post-processing step  consistently improves the AUD performance. Presumably, it refines the OOD phone labels by leveraging contextual information in the Mboshi acoustic data.




     

\subsection{Stage 2: speech segment representation and clustering}
\label{subsec:approach_2nd_stage}
This stage applies $k$-means clustering to the subword-discriminative feature representation learned by the first stage to obtain a finite set of clusters, each of which resembles a phone-like acoustic unit. The discovered units are the outcome of the proposed two-stage approach.  

Speech clustering can be realized at the segment level \cite{LeeSoongJuang} or at the frame level   \cite{chen2015parallel}.  
For segment-level clustering, we need the phone segment boundaries, and the segments need to be  represented as fixed-dimension vectors. In order to obtain the segment boundaries,  we rely on the OOD ASR system (see Section \ref{subsec:approach_1st}):  after decoding the  target speech data, phone boundary information is obtained by finding discontinuities of the frame-level OOD phone labels. This phone boundary estimation method is similar to \cite{feng2016exploit}, except that here we are using a multilingual and  IPA symbol-based OOD ASR system.

This study compares two methods to obtain the fixed-dimensional segment representation. The first is a \textbf{downsampling}   method \cite{levin2013fixed} as suggested by \cite{kamper2017embeded,Bhati2019unsupervised}: a variable-length speech segment is cut into a fixed number ($s$) of consecutive sub-segments, and 
the averages over $d$-dimensional frame-level feature vectors within each sub-segment are concatenated 
to form a  feature vector of dimension $s \times d$ for each segment.
Note that when $s=1$, the method is equivalent to an \textbf{average}-based method \cite{I3EWang} which takes the average of the frame-level features over all the frames in a segment. The downsampling method with a large $s$ captures abundant temporal information which is not captured by the average method, however a large $s$ leads to a high dimension of the segment-level feature vector which might adversely affect $k$-means. 

Segment-level clustering with the two methods mentioned above are compared with a frame-level clustering system as a baseline, which applies  $k$-means (same as  in the proposed systems)
to a frame-level feature representation. While circumventing the need for segment boundaries and the need for fixed-length segment representations, frame-level clustering  tends to produce over-fragmented discovered units \cite{wu2018optimizing,feng2019_TASLP}.  
Finally, we report an ``upperbound'' segment-level system by assuming the availability of golden phone  boundary information while keeping the other settings unchanged. This allows us to quantify the performance degradation attributed to imperfect  phone boundary estimation.


\section{Experimental setup}
\label{sec:setup}
\subsection{Evaluation metrics}
\label{subsec:setup_eval}
We use two common metrics in the AUD task \cite{ondel2016variational,Ebbers2017,Yusuf2020hierarchical} to compare with past studies: 
normalized mutual information (\textbf{NMI})  and \textbf{F-score}. NMI   measures the statistical dependency between discovered units (DUs) and ground-truth phone units (GUs), which is computed based on a frame-level confusion matrix of DU and GU  (see \cite{Yusuf2020hierarchical} for details). An NMI value ranges between $0$  and $100\%$, with a higher value  indicating a higher consistency between DUs and GUs, hence is preferred. F-score is  the harmonic mean of recall (R) and precision (P). It is used to measure  the accuracy of the phone segmentation. 
A tolerance of $\pm20$ ms is set when computing F-score values. 
Higher F-score, R and P   are preferred. 

\subsection{Databases}

\label{subsec:setup_database}
The AUD  performance    is evaluated on a corpus
 containing $5,130$ sentences spoken by three speakers of the Mboshi language  \cite{Godard2018mboshi}, 
for a total amount of $4.5$ hours. Automatically generated Mboshi phone alignments are available, but are not used during   system development. Note that the DNN AM in our system is trained and evaluated on the entire Mboshi dataset without training-test partition, as we are tackling an unsupervised learning problem. This  is also consistent with  the  studies   \cite{Yusuf2020hierarchical,Ondel2019Bayesian}.

Speech from $13$ phonetically diverse languages \cite{zelasko2020sounds} is used to train the OOD multi-/monolingual ASR systems. $5$ languages are from GlobalPhone  \cite{schultz2002globalphone}: Czech ($24$ h), French ($23$ h), Spanish ($12$ h), Mandarin ($15$ h) and Thai ($23$ h). The other $8$ languages are from IARPA Babel: Cantonese ($127$ h), Bengali ($55$ h), Vietnamese ($78$ h), Lao ($59$ h), Zulu ($54$ h), Amharic ($39$ h), Javanese ($41$ h) and Georgian ($45$ h).


 

\subsection{Implementation of stage 1}
\label{subsec:setup_stage1}
The APC model included in the first stage of the proposed model is taken from our previous study \cite{feng2020unsupervised}: 
it has $5$ LSTM layers of dimension $100$ with residual connections. The prediction step is $5$. 
The model was trained with the Libri-light (English) \textit{unlab-600 (hour)} set \cite{kahn2019librilight}.
APC features are extracted from the top layer of the APC model.


We developed $2$ multilingual systems and $5$ monolingual systems, differing  only in the training   languages: \textbf{Multi-5} denotes the multilingual system trained with the 5 GlobalPhone languages; \textbf{Multi-13} denotes the multilingual system trained with all 13 GlobalPhone+Babel languages; \textbf{Mono-CZ, Mono-FR, Mono-SP, Mono-MA, Mono-TH} are five monolingual systems, each trained with one GlobalPhone language, i.e., Czech, French, Spanish, Mandarin, and Thai, respectively.
We do not report any monolingual system trained on each of the Babel languages because of its inferior AUD performance -- possibly explained by the large recording condition mismatch  between the Babel languages and Mboshi.
IPA symbols are used to represent the basic acoustic units, and the mapping from  orthographic transcriptions  to IPA symbol sequences is obtained by LanguageNet G2P models \cite{hasegawa2020grapheme}.

The multilingual and monolingual OOD ASR systems are trained using Kaldi \cite{povey2011kaldi}, adopting a hybrid  architecture \cite{dahl2011context}, following implementation   in \cite{feng2021how}. The AM adopts a factorized time-delay neural network (TDNNF) consisting of $12$ layers with a hidden dimension of $1024$
and Resnet-style skip connections, trained with the LF-MMI criterion \cite{povey2016purely} for $4$ epochs, with a starting learning rate (LR) of $10^{-3}$. The input features consist of $43$-dimension high-resolution MFCC+pitch features and $100$-dimension i-vectors. The language model (LM) is a uni-gram phonotactic LM instead of an RNNLM, as we intend the OOD ASR phone labeling process to be  minimally affected by the OOD language phonotactics. The LM is trained with the training data transcripts using SRILM \cite{Stolcke02srilm--}.


The DNN AM for Mboshi is trained using either APC features or MFCC features of the Mboshi data. 
For each of  the $7$ multi/monolingual OOD ASR systems, the generated and post-processed (see Section \ref{subsec:approach_1st}) OOD phone labels are used    to train one DNN AM with MFCC as input features, resulting in $7$ DNN AMs. 
For the sake of simplicity, to test the effectiveness of the APC features,  only for the system employing \textbf{Multi-13} (the best-performing OOD ASR),  an additional DNN AM is trained with APC features instead of MFCCs. 
The Mboshi DNN AM adopts a  TDNNF structure similar to that of the OOD ASR AM, except: a $40$-dimension bottleneck layer is  placed below the top TDNNF layer; i-vector input is not included as we found it deteriorated the performance in our preliminary results (not included in this paper);
the model is trained for $20$ epochs with a smaller LR of $2.5\cdot 10^{-4}$ to stabilize the training procedure due to only $4.5$ hours of training material\footnote{Speed perturbation based data augmentation techniques were tried but no improvement was found, thus are not applied.}.  After training the  DNN AM, the BNF for the Mboshi representations are extracted from the bottleneck layer  as the  learned frame-level subword-discriminative representation, and are used as input to the 
second stage of the proposed system.



\subsection{Implementation of stage 2}
\label{subsec:setup_stage2}

The $k$-means algorithm is implemented using \cite{scikit-learn}. 
Unless specified differently, the number of clusters is empirically set to $50$.
Segment-level clustering is done on all the learned subword-discriminative representations of the different DNN AMs  (i.e., 1 for each OOD system). 
For segment-level clustering, the speech segment boundaries are estimated using the OOD ASR system in the first stage. 
The downsampling method with $s$ in $\{2,3,4,5\}$ and the average based method are compared for obtaining the fixed-dimension segment  representation. 
The frame-level clustering baseline and the upperbound segment-level system are based on the feature representation  learned using \textbf{Multi-13}. Since the optimal setting of the number of clusters is unknown, we tested a range between $30$ and $70$. 




\section{Results and discussion}
\label{sec:results}
For each experiment, we repeat $k$-means clustering $5$ times with different  random initialization and report NMI and F-score  in means $\pm$ standard deviation.

\subsection{Evaluation of stage 1: Effect of frame-level subword-discriminative feature representations}
\label{subsec:multi_mono_asr}
\begin{table}[!t]
\renewcommand\arraystretch{0.4}
\centering
\caption{Comparison of adopting a multi-/monolingual OOD ASR system  in stage 1 of our approach  and SotA \cite{Ondel2019Bayesian,Yusuf2020hierarchical}.}
\resizebox{ 0.83 \linewidth}{!}{%
\begin{tabular}{clcc}      
\toprule
Input &   system & NMI ($\%$) & F-score ($\%$)\\
\midrule
\multirow{7}{*}{MFCC}
& Mono-CZ&$40.87\pm0.14$ &$63.01\pm0.06$\\
& Mono-FR&$38.32\pm0.17$ &$\bm{64.14\pm 0.10}$\\
& Mono-SP&$37.57\pm0.51$ &$58.87\pm0.08$\\
& Mono-MA&$38.85\pm0.24$ &$61.45\pm0.12$\\
& Mono-TH&$37.61\pm0.08$ &$61.79\pm0.05$\\
& Multi-5 &$41.93\pm0.28$ &$62.84\pm0.03$\\
& Multi-13 &$\bm{43.00\pm 0.12}$ &$62.89\pm0.07$\\
\midrule
APC& Multi-13 &$42.15\pm0.28$ &$62.90\pm 0.15$ \\
\midrule
\midrule
\multirow{2}{*}{N/A}
&Yusuf et al.
\cite{Yusuf2020hierarchical}& $41.07\pm 1.09$ & $59.15\pm 1.51$ \\
&Ondel et al. \cite{Ondel2019Bayesian}& $38.38\pm0.97$ & $59.50\pm 0.78$ \\
\bottomrule
\end{tabular}%
}
\label{tab:results_mono_multi}
\end{table}

We first evaluate the effect of using a multilingual vs. a monolingual OOD ASR system on the effectiveness of stage 1. Next, we investigate the effect of using APC features as input features to the DNN AM. A fixed setting  of stage 2 is used in all these experiments, i.e., the segment-level $k$-means with the average-based method to obtain the segment-level feature representation. The performances of our systems and two state-of-the-art (SotA) systems from the literature \cite{Yusuf2020hierarchical,Ondel2019Bayesian} are listed in Table \ref{tab:results_mono_multi}. Several observations can be made from this table:

(1) The multilingual OOD ASR systems (Multi-13/Multi-5) outperform the monolingual systems on the NMI measure. Apparently, using a language-independent OOD ASR system to provide the OOD phone labels is better than using a language-dependent system for target language DNN AM training, to learn a better frame-level subword-discriminative feature representation. Looking at the F-score, the Multi-5 and Multi-13 systems perform   better than the average over the $5$ monolingual systems ($61.85\%$), nevertheless Mono-FR achieves the best performance, followed by Mono-CZ.  
The results indicate that using a multilingual OOD ASR  is more beneficial for improving the phonetic relevance of the discovered acoustic units (NMI) than for improving phone boundary estimation (F-score).  

(2) Our best system (Multi-13 with MFCC input) outperforms state-of-the-art \cite{Yusuf2020hierarchical,Ondel2019Bayesian} on both NMI and F-score.  Similar to our approach,    \cite{Yusuf2020hierarchical,Ondel2019Bayesian} also relied on OOD languages' transcribed data for model training. However, they used around
$35$ hours (from $7$ languages), while our two best systems (in NMI) used more OOD speech data - Multi-13: $595$ hours; Multi-5: $97$ hours. Nevertheless, our  Mono-CZ system performs on par with or better than \cite{Yusuf2020hierarchical,Ondel2019Bayesian} in NMI and F-score respectively, while using only $24$ hours of Czech data.

(3) Comparison of the two Multi-13 systems shows that  APC features as input to the DNN AM in stage 1 does not affect the F-score and deteriorates the NMI performance compared to MFCCs. This result, seemingly in contrast to \cite{feng2020unsupervised}, can be explained by the following:
in \cite{feng2020unsupervised}, the APC model was trained on the target language English, while here the English-trained APC model was used to capture the target language (Mboshi). Moreover, \cite{feng2020unsupervised} showed the success of APC features for the USM task, here we used a different, AUD task.


\subsection{Evaluation of stage 2: Clustering strategies}
\label{subsec:clustering_strategies}
\begin{table}[!t]
\renewcommand\arraystretch{0.4}
\centering
\caption{Speech clustering strategies in stage 2 of the proposed approach. ``Seg./Fra.'' denotes segment- and frame-level clustering.
``AVG$^\ddagger$'' indicates the upperbound system.  }
\resizebox{ 1 \linewidth}{!}{%
\begin{tabular}{llcccc}      
\toprule
Type &System & NMI ($\%$) & F-score ($\%$) & \textoverline{Recall} ($\%$)& \textoverline{Precision} ($\%$) \\
\midrule
\multirow{6}{*}{Seg.}
&AVG &$\bm{43.00\pm 0.12}$ &$\bm{62.89\pm0.07}$ & $73.47$ & $54.97$ \\
&DS-2 &$\bm{43.00\pm 0.12}$ &$62.87\pm0.07$& $74.22$ & $54.54$\\
&DS-3 & $42.73\pm0.28$ & $62.70\pm0.15$ &$74.12$& $54.32$\\
&DS-4 & $42.49\pm0.27$& $62.47\pm0.10$ &$73.74$&$54.19$\\
&DS-5 & $42.44\pm0.16$& $62.60\pm0.07$ &$74.07$&$54.20$\\
\cmidrule(lr{1em}){2-6}
&AVG$^\ddagger$ &{\color{black}$59.29\pm1.17$}& {\color{black}$97.73\pm0.06$} &{\color{black}$100.00$}  & {\color{black}$100.00$}\\
\midrule
Fra. & Baseline & $41.82\pm0.20$& $43.59\pm0.35$ & $90.38$  & $28.72$  \\
\bottomrule
\end{tabular}%
}
\label{tab:results_segment_strategies}
\end{table}
\begin{table}[!t]
\renewcommand\arraystretch{0.4}
\centering
\caption{NMI (row 2) and F-score (row 3) performances w.r.t different numbers of clusters (row 1).}
\resizebox{ 0.99 \linewidth}{!}{%
\begin{tabular}{ccccc}      
\toprule
$30$ & $40$ & $50$ & $60$ & $70$ \\
\midrule
$41.35\pm0.21$& $42.50\pm0.34$&$43.00\pm0.12$ & $\bm{43.22\pm0.52}$&$43.20\pm0.53$\\
\midrule
$62.81\pm0.05$&$\bm{62.89\pm0.04}$ &$\bm{62.89\pm0.07}$ & $62.77\pm0.14$&$62.76\pm0.10$\\
\bottomrule
\end{tabular}%
}
\label{tab:results_num_clusters}
\end{table}

We investigated the effect of the two segment representation strategies in stage 2, i.e., the downsampling method with different $s$ and the average-based method, and compared these to the frame-level baseline and the system with access to golden phone boundary information (upperbound: ``AVG$^\ddagger$''). The Multi-13 OOD ASR system is used to generate the OOD phone labels in all experiments; APC is not adopted. 
Table \ref{tab:results_segment_strategies} shows the results. ``AVG'' denotes the average-based method, ``DS-2$\sim$5'' denotes the downsampling method with $s=2\sim5$.
In addition to NMI and F-score, Table \ref{tab:results_segment_strategies}   reports  the average  \textit{recall} and \textit{precision} values for each system, in order to gain deeper insights into the differences between segment- and frame-level $k$-means.

It can be clearly seen that the systems adopting segment-level $k$-means outperform the frame-level baseline  on NMI and F-score. The superiority of the segment-level systems is more prominent on the F-score (absolute $19.0\%$)  than on NMI (absolute $1.2\%$). Particularly, the baseline model has a very low \textit{precision}, indicating a large proportion of false boundaries  that are hypothesized. 
This implies a frame-level system tends to over-segment target speech, which is in line with   \cite{wu2018optimizing,feng2019_TASLP}.  

Table \ref{tab:results_segment_strategies}  shows that the downsampling method does not have an advantage over a simpler, average based method, and a larger $s$ leads to a slight NMI degradation. While other studies showed a good performance for the downsampling method \cite{kamper2017embeded,Bhati2019unsupervised}, we show that in  this low-resource Mboshi   database, using the $k$-means algorithm,  the average based method is comparable, if not better than the downsampling method.

Finally, Table \ref{tab:results_segment_strategies} shows that the upperbound system (AVG$^\ddagger$) outperformed our best system (AVG) by $16.3\%$ absolute NMI. 
The   NMI gap is attributed exclusively to the imperfect phone boundary estimation by the OOD ASR system. It is expected that by improving the phone boundary estimation, or by adopting an interactive approach to refining segmentation and target language acoustic modeling \cite{kamper2017embeded}, the  frame-level subword-discriminative representation learned in the first stage  could be based on to achieve an NMI that approaches the upperbound. 




The effect of the number of clusters on the AUD task was investigated using the average based method.
Table \ref{tab:results_num_clusters} summarizes the results: the best NMI is obtained with a  number of clusters  between $60$ and $70$. 
The F-score performance is less sensitive to the number of clusters than NMI. 
Overall, a number of clusters between $50$ and $70$ shows the best performance.


\section{Conclusions and future work}
\label{sec:conclu}
This paper proposes a two-stage approach for the unsupervised AUD task. Our best model, which employs $13$ OOD language resources in stage 1 to provide phone labels for target language AM training and uses an average-based method segment clustering in stage 2, outperforms state-of-the-art performance on a very low-resource Mboshi database. The results showed that the multilingual OOD ASR systems outperformed a monolingual one in providing the frame labels for target language acoustic modeling and in phone boundary estimation, with the former being more prominent. Comparison with a golden standard showed that a $16.0\%$ NMI performance gap could be attributed to imperfect phone boundary information. Furthermore, in stage 1, APC features were compared to MFCCs as input to the DNN AM training module and were less effective in the AUD task. In stage 2, the best performance was achieved using a number of clusters between $50$ and $70$.



\bibliographystyle{IEEEtran}
\bibliography{mybib_short}

\end{document}